# Holographic Volumetric Additive Manufacturing

*Maria I. Álvarez-Castaño[1], Andreas Gejl Madsen[2], Jorge Madrid-Wolff[1], Antoine Boniface[1,3], Jesper Glückstad[2], Christophe Moser[1]*

[1]*Laboratory of Applied Photonics Devices, School of Engineering, Ecole Polytechnique Fédérale de Lausanne, CH-1015, Lausanne, Switzerland*
[2]*SDU Centre for Photonics Engineering, University of Southern Denmark, DK-5230 Odense M, Denmark*
[3] *current address: AMS Osram, Martigny, Switzerland*

**Abstract:**

Three-dimensional printing has revolutionized the manufacturing of volumetric components and structures in many areas. Different 3D printing technologies have been developed including light-induced techniques based on the photopolymerization of liquid resins. In particular, a recently introduced method, so-called Tomographic Volumetric Additive Manufacturing (VAM), allows the fabrication of mesoscale objects within tens of seconds without the need for support structures. This method works by projecting thousands of amplitude patterns, computed via a reverse tomography algorithm, into a resin from different angles to produce the desired three-dimensional shape when the resin reaches the polymerization threshold. To date, only amplitude modulation of the patterns has been reported. Here, we show that holographic phase modulation unlocks new capabilities for VAM printing. Specifically, first, the effective light projection efficiency is improved by at least a factor of 10 over amplitude coding: second, the resolution can reach the light diffraction limit; and importantly phase encoding allows to control ballistic photons in scattering media, which potentially increases the volume of 3D objects that can be printed in opaque and non-absorbing resins. The approach uses computer-generated holograms to convert phase, encoded on a 2D modulator to the desired intensity projections by light propagation in a photosensitive resin container. We demonstrate the potential of holographic phase coding using simulations and experiments, the latter by implementing a volumetric printer using a digital micro-mirror device (DMD), as the 2D phase modulator in a Fourier configuration. Specifically, thanks to the Lee Holograms that allow us to use the DMD as a fast phase modulator by encoding the phase in a binary hologram. Combining tiled holograms with point-spread-function shaping mitigates the speckle noise typically associated with computer-generated holograms and speed-up their computation. We use these holographic projections to fabricate millimetric 3D objects in less than a minute with a resolution down to 164 µm.

## I.    Introduction

The fabrication of objects with complex geometry has become much simpler thanks to additive manufacturing. Light-based 3D printing exploits the ability of certain light-sensitive molecules to trigger polymerization or crosslinking reactions in liquid resins, thus solidifying them. Photopolymerization can be produced sequentially, as in stereolithography [1] or digital light processing [2], [3], and more recently also volumetrically, as in two-photon polymerization [4], [5], light-sheet microprinting [6], or tomographic volumetric additive manufacturing [7], [8]. By printing in a volumetric fashion, support struts are no longer needed, enabling the fabrication of designs with cavities and overhangs.

In Tomographic Volumetric Additive Manufacturing (TVAM), an entire three-dimensional object is simultaneously solidified by irradiating a volume of photocurable liquid resin from multiple angles with dynamic light patterns [7], [9]. Unlike most other additive manufacturing methods, tomographic volumetric additive manufacturing is layer-less, meaning that it does not fabricate objects by solidifying one voxel, one line, or one layer at a time. Instead, light from subsequent tomographic patterns builds up an energy dose within the complete volume of the target object. Typical printing times are tens of seconds [10] for cm-scale prints with resolutions down to 50 to 80 µm [8], [11] using high power laser diodes [8], [12], [13] . Additionally, the technique has proven versatile and has been used to fabricate objects in materials such as acrylates, thiol-enes [11], [14] [15], [16], nanoparticle-loaded composites [11], polymer-derived ceramics [17], epoxies [18],  silk bioinks [19], and cell-laden hydrogels [9], [20], [21], [22], [23].

So far, implementations of tomographic VAM rely on the Radon transform to calculate the light amplitude patterns. In essence, due to the positive intensity patterns constraint and the resin's hardening properties, the computed projection patterns produce a 3D dose with artifacts resulting in printing objects with poor fidelity. The computed patterns may be iteratively optimized [7], [24], using feedback from the printing process [8] [25], [26], or corrected for scattering [12] and by incorporating chemical diffusion [27].

The Radon transform assumes straight rays to compute the projection light patterns. Projection patterns can be thought of as an "extruded" two-dimensional patterns that propagate as a collimated beam. In practice, this is implemented by imaging a two-dimensional spatial light modulator into a cylindrical vial with the resin and relying on the depth of focus of the imaging system to maintain a quasi-collimated image. If the vial's diameter is much larger than the depth of focus, blurring artifacts appear and compromise resolution[8], [13]. The étendue of the light source is a key optical parameter here, as it limits for how long a high-resolution image can be kept collimated. Additionally, in this type of optical configuration, the two-dimensional spatial light modulator modulates intensity, and not phase. Thus, one pixel on the 2D modulator corresponds to one pixel in the relayed image. Typically, digital micromirror devices (DMD) are used as reflective binary amplitude modulators in light-based 3D printers including TVAM. This microelectromechanical system consists of a 2D array of thousands of microscopic mirrors that modulate the light by tilting the mirrors between ± 12° corresponding to two states ON or OFF, and modulates greyscales the mirrors switched quickly by binary pulse-width modulation.

Alternatively to amplitude, phase can be modulated to include additional information into the light patterns. Phase encoding of tomographic projections offers multiple advantages over amplitude modulation. First, phase-encoding improves light efficiency, as all pixels of the display can contribute to the projected intensity pattern. Instead, when using amplitude projections, energy is wasted as most of the pixels are off in the tomographic patterns. Secondly, phase-encoding allows the projection control of beams to remain in focus for longer extents, reducing the disparity between Radon-based calculations and the actual projections. This can be done by projecting a sequence of diffractive phases that are in focus at different depths, effectively producing a low-divergence projection over time and smearing the intensity reconstruction over the build volume. Thirdly, phase-encoding uses a simpler optical setup, that may be also tailored to correct for down-stream optical aberrations. In this work, we use a projection system that converts a two-dimensional phase modulator to a two-dimensional intensity pattern in the Fourier plane. In this configuration, light reflected by one pixel on the modulator is spread to a large collimated beam in the optical Fourier plane. The angle at which light is spread depends on the position of the input pixels. Remarkably, all pixels on the modulator contribute, by holographic self-interference, to all pixels on the image plane. There are multiple iterative algorithms or machine learning approaches [28] that can generate two-dimensional input phase patterns to produce a user-specified intensity distribution in a desired plane or volume. The Gerchberg–Saxton (GS) iterative algorithm is one of the many methods to calculate a two-dimensional input phase pattern, which results in a desired intensity distribution pattern at the optical Fourier plane [29]. Binary holograms computed with the Gerchberg-Saxton algorithm and encoded with Lee holograms have recently been used to control multiple foci in two-photon lithography, thereby increasing printing speed [30].

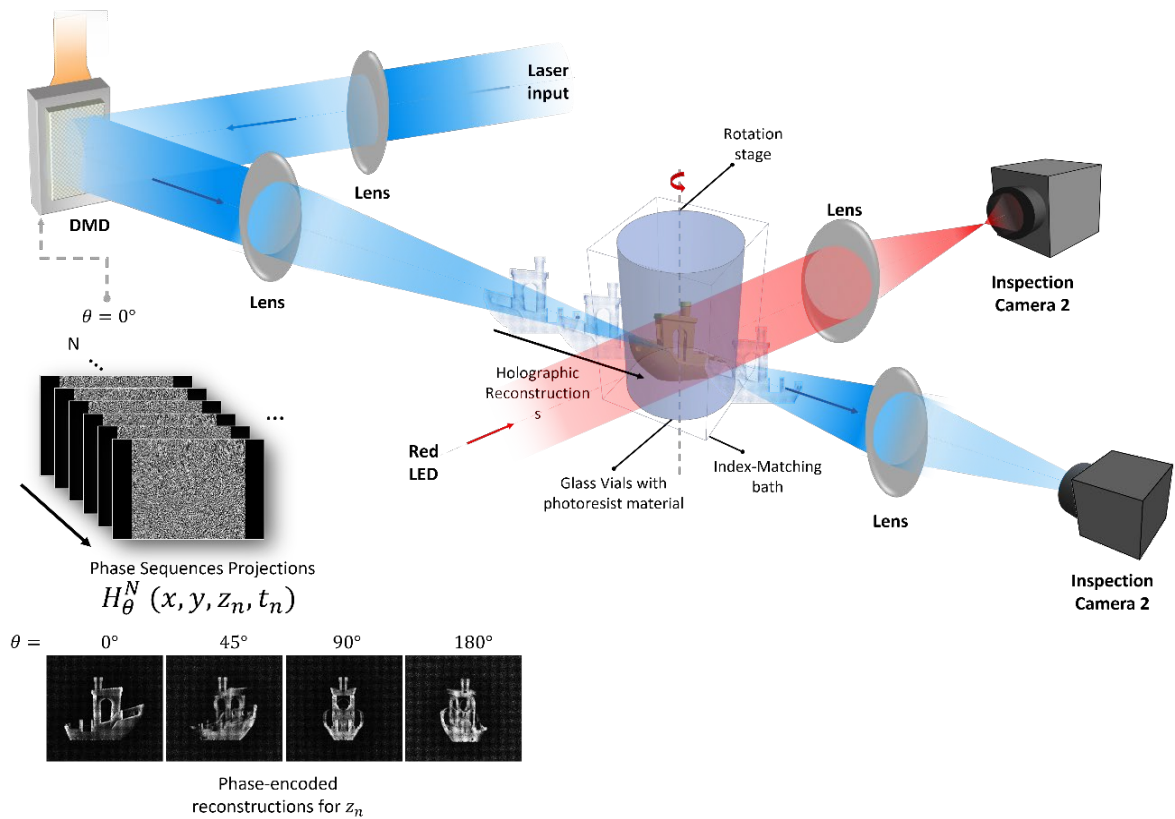

**Fig.1.** Experimental setup for holographic volumetric additive manufacturing.

CGHs typically suffer from speckle noise, which generates grainy images [31], [32] [33], [34]. In the context of VAM, speckle noise needs to be avoided in the projected intensity image because they could create unwanted printed speckled voxels. To improve the reconstruction quality of the phase-only holograms, several methods such as time averaging iterative algorithms applying bandwidth constraints [35], [36], [37], and or camera in the loop [38] have been proposed. Due to the uncontrollable interference between adjacent pixels in a CGH caused by overlapping Airy disks, tiling holograms in space and/or time or by other output pixel separation methods can provide attenuation of unwanted noisy interference [39], [40], [41], [42].

Here we combine time-multiplexing [43] and tiled holograms [39], [40] using the HoloTile method [42], [44] to produce near speckle-free projections, as will be discussed in section III. In general, the computational time for full-size CGHs scales as $MlogM$, where $M^2$ is the number of pixels in the image. An additional advantage of tiled holograms is the reduction of the computation time by at least an order of magnitude due to the fact that the dimensions of the tiles (sub-holograms) are smaller than those of a full-size hologram and only the smaller sub-holograms need to be computed.

Moreover, holographic projections give us the possibility to engineer the shape of the projected output "voxel shape" in 3D thanks to wavefront control.

Spatial Light Modulators (SLM) based on liquid crystals are typically used to modulate complex light fields; but they are not well suited for VAM due to their low frame rate (20-100 Hz) and reduced resistance to short wavelengths (< 450 nm). Thus, we use a DMD, which is a binary amplitude SLM. To unlock the phase modulation of the DMD we implement the binary Lee hologram method [45], [46]. The synthesis of diffractive phase encoding using Lee holograms has been of increasing interest in various applications thanks to the high phase modulation rate of DMDs [30], [32], [46], [47].

Phase encoding requires coherent light both spatially and temporally. This means that single mode lasers have to be used. A single mode laser has less optical power than a multimode laser and thus power efficiency is of high importance to obtain enough intensity at the build volume, keeping printing times short. For VAM, the parameter that defines resolution is the étendue of the light source (product of the source surface by the emission solid angle). The lower the étendue, the more light can be collimated and therefore approximate the Radon transform. For example, a 405 nm LED with an emission area of 1 mm$^2$ and emission angle of 170° has an étendue which is $10^6$ times larger than a single mode 405 nm laser diode (1 µm x 2 µm emission area and 30° by 15°). In essence, the divergence of the light beam yields a spatial resolution which varies throughout the volume[8], [13]. In TVAM, the best lateral resolution is obtained when a single-mode laser is used.

The optical configuration is presented in Fig 1. The system consists of a continuous wave (CW) laser diode at 405 nm, collimated and expanded to fit the active area of a DMD. Unlike conventional tomographic VAM [7], [8], the DMD is arranged in a so-called *2f* optical Fourier configuration, which allows the reconstruction of the projected holograms into the printing volume. The resin container is set to rotate at constant speed by a mechanical stage while the holographic projections are displayed synchronously. A sequence of *N* binary phase patterns $\varphi_\theta^N$ are displayed at each projection angle to smear the holographic projection in the building volume. Binary phase encoding of the DMD is performed at a rate of $\sim 1.6\ kHz$, while the stage rotates continuously at a speed of 30°/s. We discretize the rotation into angles of $\Delta\theta = 0.6\ °$. For each of these angles, we typically calculate 32 patterns, each focused at a different depth. All patterns for each $\Delta\theta$ are sequentially projected within 20 ms. The short projections time should not blur the inverse Radon projections at the angular speed of $30°/s$. Two inspections cameras are used to monitor the holographic pattern reconstruction (camera 1) and the polymerization process (camera 2).

Here, we demonstrate the use of phase encoding in tomographic VAM by printing a millimeter-scale object in an acrylate-based resin. We demonstrate a printing time of 60 s with a single cw 40 mW UV laser diode. This short printing time is the result of the efficiency of the light engine: first, we show a comparison between the light efficiency for amplitude encoding and phase encoding. We then present a pipeline to compute holograms for tomographic projections, where tiled holograms reduce speckle noise and increase contrast. Fresnel lenses were additionally implemented to smear the projection along the optical path, synthetically reducing the etendue. The Lee hologram method is the last stage of the phase encoding that allows to use the DMD as a fast phase modulator. The threshold in the polymerization kinetics of the resin was analyzed in relation to speckle noise.

## II. Light engine efficiency: amplitude encoding

In tomographic printing, the Radon transform $R(r,\theta)$ is digitally computed to produce angular projections followed by the Filtered back-projections (FBP) [48]. The process to synthesize the amplitude projections is shown in Fig. 2. First, the 3D model (typically as .stl file) is voxelized and sliced. The $y$ axis is selected parallel to the rotation axis of the vial (Fig. 2B). The Radon transform is applied to each slice to produce the sinograms $R(r,y,\theta)$ (Fig.2C). The Fourier transform of the sinogram at each angle is applied followed by a Ram-Lak linear filter. The filtered sinograms include negative values, which can't be projected. All negative values are thus set to zero in a non-negativity constraint (Figure 2D (optimized projections)). Projections were calculated over 360° with angular resolution of 0.6°.

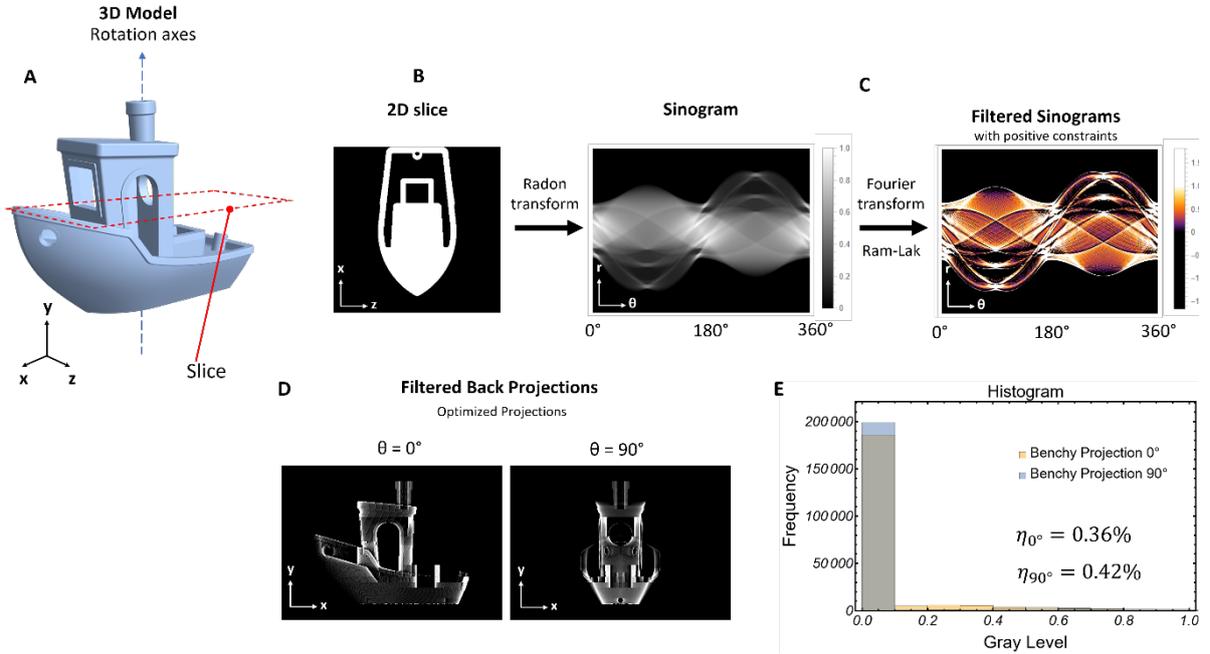

**Fig. 2.** (**A**) 3D target object. The dashed lines represent a plane slicing the object perpendicular to the direction of rotation. (**B**) 2D slices from the 3D object. (**C**) Sinograms of each layer after applying the Radon Transform. (**D**) Filtered back projections after applying a Ram-Lak filter and setting negatives values to zero. (**E**) Histograms of two sparse projection patterns at 0 and 90 degrees.

$$R(r,\theta) = \int_{-\infty}^{\infty}\int_{-\infty}^{\infty} f(x,z)\delta(x\cos\theta + z\sin\theta - r)dxdz \qquad (1.0)$$

Where $\delta()$ is the Dirac delta function, $r$ represents the spatial coordinates, and $\theta$ represents the projection angle.

Typically, in VAM systems, a DMD is used as a fast amplitude display to project amplitude patterns at high rates into the resin vial. Due to the nature of the display and the sparse amplitude of tomographic projections, most of the optical power incident and reflected by the DMD is lost, with only a few pixels contributing to the projected image. This is exemplified in the histograms in Fig. 2E, where each pattern shows a large amount of dark pixels. An estimate of light efficiency of each pattern is performed using the following relation:

$$\eta_{patt} = \frac{\sum_{g=0}^{255} n_g \frac{g}{255}}{N_{pixels}} * \eta_{DMD} \qquad (2.0)$$

Where $n_g$ is the number of pixels in the image having a gray level equal to $g$. The maximum gray level is 255 because of the 8-bit encoding of the DMD (via pulse width modulation). $N_{pixels}$ is the number of pixels of the image, and $\eta_{DMD}$ is the inherent efficiency of the DMD at the operating wavelength. As an example, the histogram analysis using theoretical projections images of the Benchy boat at projection angles $\theta = 0°$ and $\theta = 90°$ shows a light efficiency $\eta_{patt}$ for amplitude modulation of 0.36 % and 0.42 % respectively (assuming a pixel reflectivity $\eta_{DMD} = 65\% \: at \: 405 \: nm$). Experimentally, the light efficiency has been measured slightly below the theoretical value (0.35 % and 0.40 % respectively). Because the Radon projection patterns are sparse, amplitude modulation has a very poor light efficiency in VAM tomographic systems. For example, the VAM system from Loterie et al. [8] combines six high-power broad-area laser diodes that produce 1.6 W at the output of a 70 µm square core optical fiber. After amplitude modulation by the DMD [8], the light intensity is reported to be in the order of $5.4 \: mW$, corresponding to a modulation light efficiency of 0,34 %. The reason for this low light efficiency is that there is a one-to-one relation between the plane of the modulator and the plane of the image. In contrast to the above intensity-based projection, for holographic phase modulation with a CGH pattern, each pixel of the modulator contributes to the intensity pattern which is expected to yield a substantially higher light efficiency (see light efficiency measurements in the Supplementary Material).

### III. Light engine efficiency: holographic encoding

Speckle is an inherent holography problem [32]. While a CGH can reconstruct the desired intensity distribution at the Fourier plane, the reconstruction quality is affected by the CGH quality, the pixelated structure of the applied 2D display, and all the physical phenomena involved in the propagation of light, such as interference, multiple diffraction orders, and optical aberrations [39]. A typical intensity reconstruction from a CGH displayed on a pixelated 2D display is shown in the Supplementary Material S-Fig.2 A-D, and the reconstructions are affected by speckle noise. To improve the reconstruction quality of phase-only holograms, a plurality of methods have been proposed such as using time averaging [43], [49], and iterative algorithms applying bandwidth constraints [35], [36], [37].
Spatial and/or temporal tiling of holograms can be used to mitigate the effect of unwanted crosstalk between pixels [39], [40], [41], [42] i.e., Fig. 3A. The applied HoloTile modality, described in detail in Supplementary Material S.6, consists of superimposing spatially tiled holograms with a phase front to modify the point spread function (PSF) of each reconstructed pixel, which helps to increase the contrast of the reconstruction according to the desired PSF, i.e., Fig. 3B-C. The key aspect of tiling hologram is to provide high-speed computation, and separate reconstructed image points (modifying the original grid of image points by changing the size of the tiles), which eliminates unwanted interferences [39], [40], [42], the PSF shaping modulates the output pixel size i.e., Fig. 3B-C.

The space between grid points in the reconstructed plane provides space to shape the PSF at will. It can be a shifted focused point, a flat top, a Bessel beam for example. The gray level value of each grid point can be specified independently. An extended Bessel beam PSF would provide a longer collimation of the projected image than with a flat top type PSF, which could be very interesting to implement for a Radon smeared image for example. A strategy to increase resolution could be by multiplexing the tiled holograms in both space and time while PSF shaping tiled holograms by adding shifting phase masks to avoid the loss of resolution but still preserve the overall fidelity [50]. Fig. 3D shows the how the grid can address different points by tiling in a different manner over time and shifting the phases.

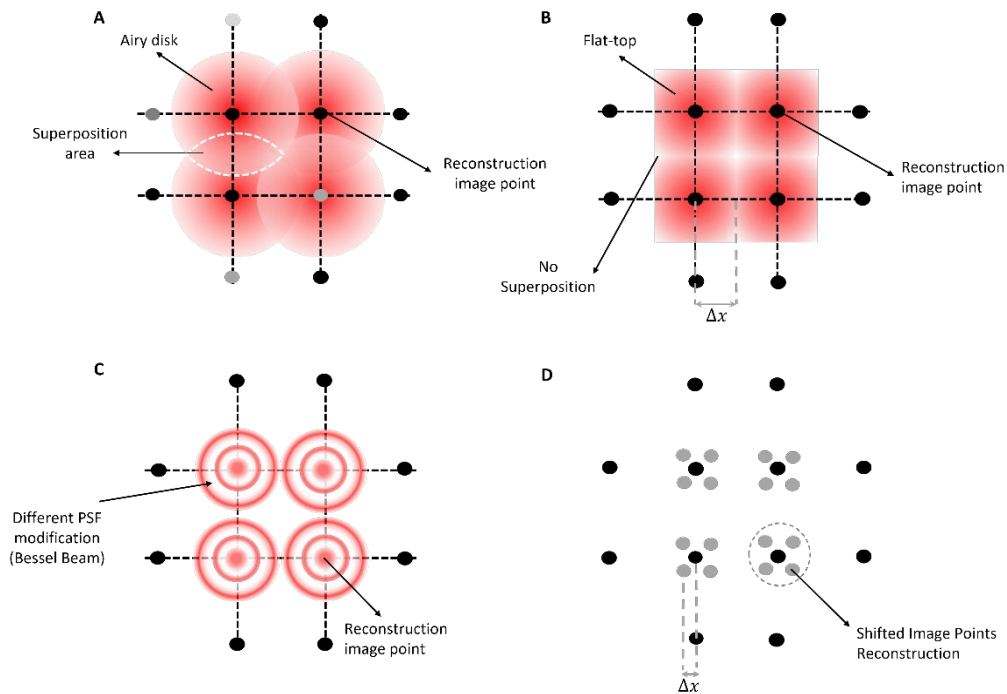

**Fig 3**. (**A**) Image points illustration with an Airy pattern intensity distribution, showing unwanted interference in an overlap area. (**B**) Image points illustration with a PSF modification corresponding to a flat-top, Δx is the lateral spacing between frequency components. As there is no overlap between the image points, speckle noise is reduced in the resulting projections. (**C**) Illustrates the flexibility of the method by shown the image points reconstruction with a different PSF modification just modulates the output pixels. (**D**) Illustrates the modification of the grid of the image points possible by changing the tiling in time or using phase shifting.

The number of tiles $N_t$ determines the dimensions of the sub-holograms $m \times m = \frac{L}{N_t} \times \frac{L}{N_t}$. Where $L \times L$ is the dimension of a full hologram size (the full size of the SLM display, in pixels). The sub-hologram $\varphi_{SubH}(x, y)$ is calculated with the GS algorithm (see Supplementary Material), but this time the hologram has a size of $m \times m$ which is much smaller than the full, size of the SLM and thus this reduces the computational burden by a factor $MLogM / mLogm$. The sub-hologram are tiled in a $L \times L$ square hologram $\varphi_{tile}(x, y)$, which is a mosaic of $N_t^2$ sub-holograms $h(x, y)$, see Fig 5A stage I.To experimentally investigate the effect of tiling the holograms on the reconstruction quality, six CGHs with the same target ("A" letter) but tiled with different $N_t$ were generated. We used a flat top PSF shaping function. The results of the reconstructions are shown in Fig. 4A. To compare changes in the reconstruction, the contrast is quantified with the mean square error (MSE), and the peak signal-to-noise (PSNR) versus the number of tiles. Results are shown in Fig. 4B-D. As seen in Fig. 4B-C, MSE decreases and PSNR increases with the number of tiles, indicating speckle-noise reduction. Although the PSNR increases with the tiles, the values are low. This might be a consequence of the flat-top shapes. A low MSE and increasing PSNR indicates a low level of speckle noise. Multiplexed holograms decrease the MSE and increase the PSNR even more. We define the contrast over the region defined by the letter "A" as a "goodness" metric for the fidelity of the reconstruction. The higher the contrast, the better the fidelity. As the number of tiles increases, the contrast of the reconstruction increases (Fig. 4D), while the resolution decreases.

The number of tiles and the sampling rate $\Delta x$ provides a relation the number of controllable grid points in the reconstruction plane and the feature size respectively [39], [40], [42], [50]. Because of the use of a flat top PSF, the minimal feature size in the projections (Fig 4E) is distance between the grid in the reconstruction plane. Smaller feature size can be realized with different PSF. The smallest being a focus grid point shifted by a distance equal to $\lambda/NA$, where NA is the numerical aperture of the Fourier Lens or projections lens (PL).

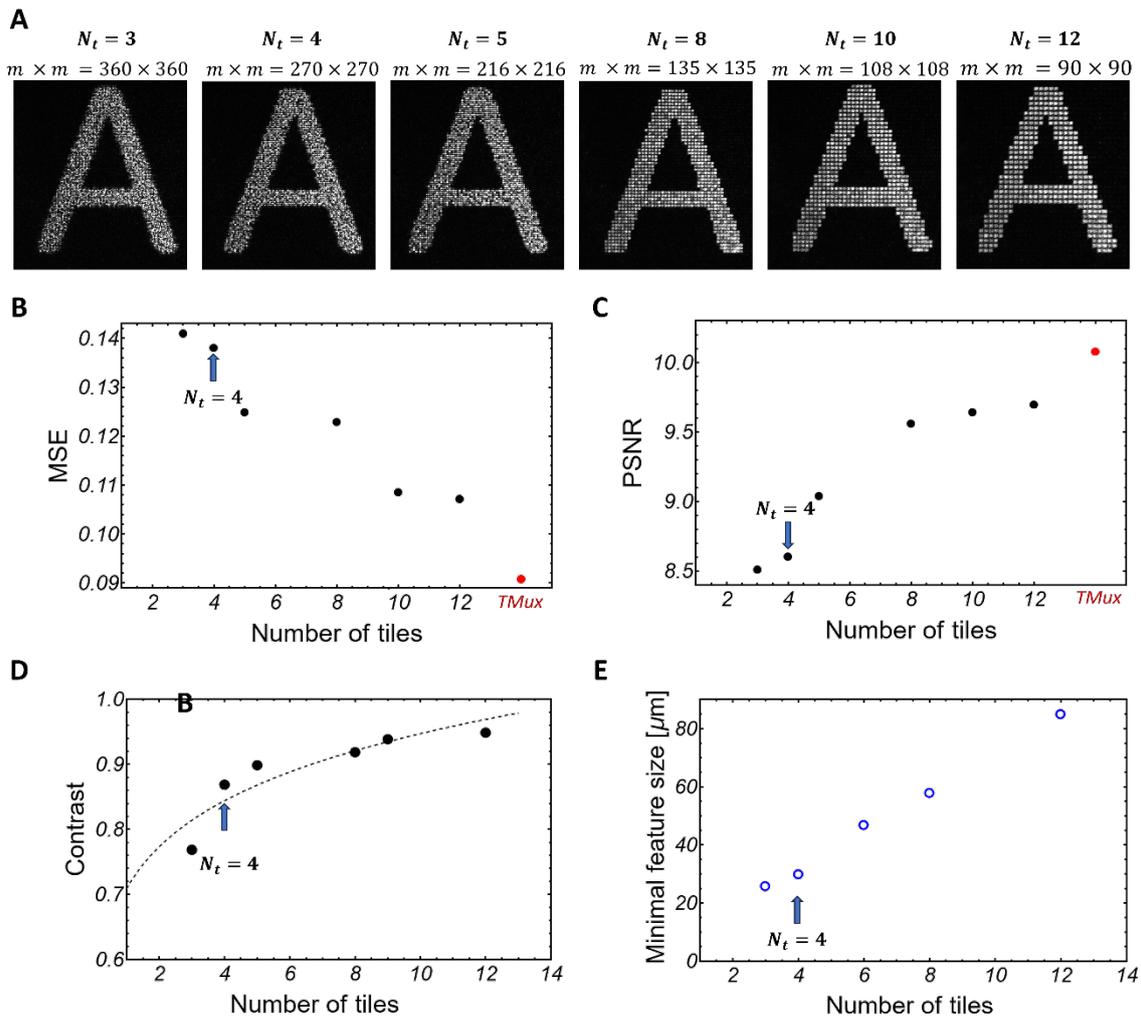

Fig 4. **Tiled holograms evaluation.** (**A**) Experimental reconstruction of the tiled holograms generated of the letter "A" for a different number of tiles $N_t$. The experiments were performed using a liquid crystal SLM (See Supplementary Material). Comparison of the measured (**B**) mean square error (MSE) and (**C**) peak signal to noise ratio (PSNR) for the different tiled holograms. (**D**) Reconstruction analysis considering the contrast measurements as $(I_{max} - I_{min}/I_{max} + I_{min})$. (**E**) Minimal projected feature size measured for different tiled holograms using the DMD.

## Computation of holograms for tomographic projection

Fig. 5 shows the pipeline to compute holographic projections. A CGH with tiled holograms and PSF shaping provides speckle noise reduction at the cost of lower resolution. The phase CGHs are then converted into binary amplitude for the DMD using Lee holograms. At each angle, the projections are sequentially reconstructed at different depths $z$ in the vial, which has the effect of "smearing" the reconstruction of the projection over $z$ giving an axial control.

Fig.5B-D shows the reconstruction of a CGH corresponding to the 3D Benchy boat projection at $\theta = 0°$. The separation between the image point in the reconstruction plane for a tiled hologram with $N_t = 4$ could be seen in Fig. 5B. The separation of the points avoids the crosstalk between pixels, but also allows us to modify the PSF for a flat-top that helps us to increase the efficiency of the illumination and resolution Fig 5C. The effect of the Lee hologram method to encode the phase and the use linear carrier of frequency $\nu_0$ is shown in Fig. 5D. Here, the complex field modulation is possible with an off-axis hologram. The +1 order is selected while the -1 and the zero order are filtered out in the Fourier plane. A $4f$ system relays the +1 order onto the printing hologram.

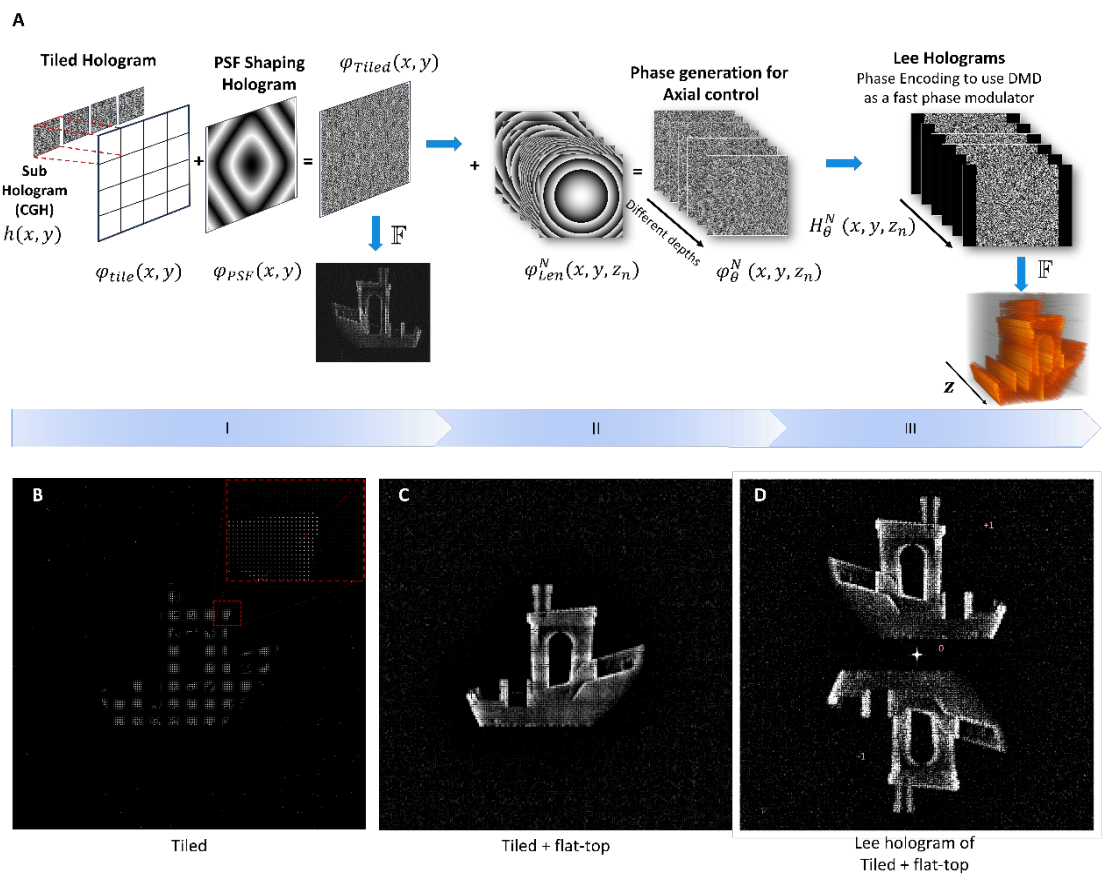

**Fig 5. Hologram synthesis. (A)** Pipeline of the hologram synthesis for holographic projection. Each stage of the process is performed for each projection. **(B)** Tiled hologram reconstruction without the superposition of the flat-top phase, **(C)** Tiled hologram reconstruction with the Flat-top phase, **(D)** Lee hologram reconstruction of the tiled phase

Next, we demonstrate the advantage of using holographic phase encoding to produce low-divergence projections. As we discussed in the introduction, low divergence is required to match the assumption of a collimated beam from the Radon transform. Fig. 6 shows the projections as they compare to the required build volume in the 12 mm vials used for printing (Fig. 6A). As seen in Fig. 7B, the recorded projections of an extruded gear remain in focus along 10 mm of the optical axis. Thanks to phase encoding, we can sequentially display multiple CGH, each focusing on a different depth along the optical axis. Simulations and recordings of these projections along an axial cross-section show that the resulting projections remain in focus throughout the vial's volume, illustrated by a dotted circumference in Fig. 6C.

An example of the incoherent sum generated by displaying a projection sequence of 25 CGHs at a 8 kHz rate is shown in Fig 6B. The incoherent sum of the projected patterns is experimentally captured by a camera on a motorized stage moving with a $20\mu m$ step while the sequence of holograms is projected by the DMD. This effectively measures the optical dose received by the resin through the volume for a given angle. The propagation distance z (10mm) is large compared with the object dimension to illustrate that the object retains the same shape along the propagation direction (Fig. 6D). This experiment shows how the shape (cogs of the gear) is preserved over propagation. Experiments match simulations results well (Fig. 6C). At each angle, the sum of ~32 patterns produce unwanted dose distribution due to diffraction. Speckle noise is minimized at the reconstruction plane but not necessarily away from it. To investigate the effect of this noise, a threshold, corresponding to the hardening point of the resin is applied.

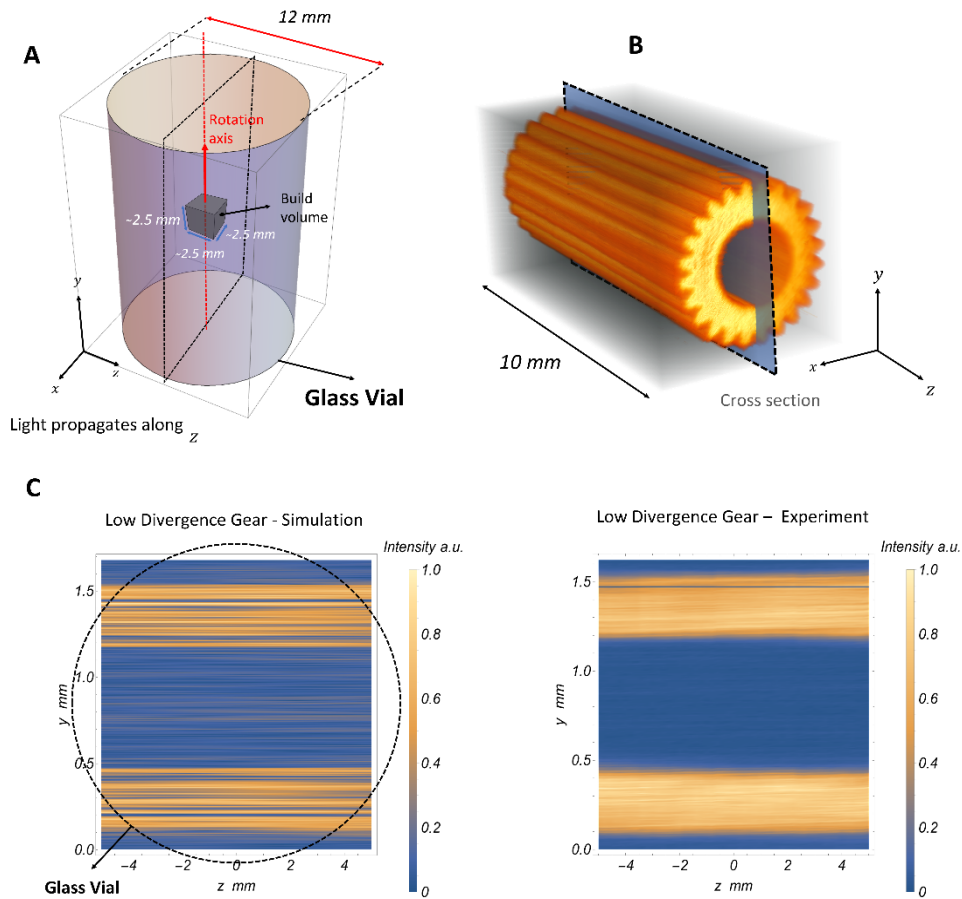

**Fig 6. Low divergence projections thanks to holographic phase encoding.** (**A**) Build volume within the glass vial. (**B**). Registered 3D rendering of cumulated light projected by a sequence of holograms. (**C**). Axial cross-section of the cumulative light dose.

As the HoloTile-based computer-generated holograms are projected into the rotating vial with photocurable resin, a 3D energy distribution is deposited, eventually leading to polymerization. Only those regions which receive sufficiently high doses of energy will polymerize [51]. Thanks to this thresholded behavior, objects can be successfully printed even if the background has a positive light intensity level, as shown in Fig. 7. In these computer simulations, we see that if the deposited dose is kept low, the object can be resolved from the background. This demonstrates that the signal-to-noise ratio of the projections is sufficiently high and does not hinder printing. If hologram projection is not stopped in time, the full build volume would polymerize and the object would not be resolved from the background.

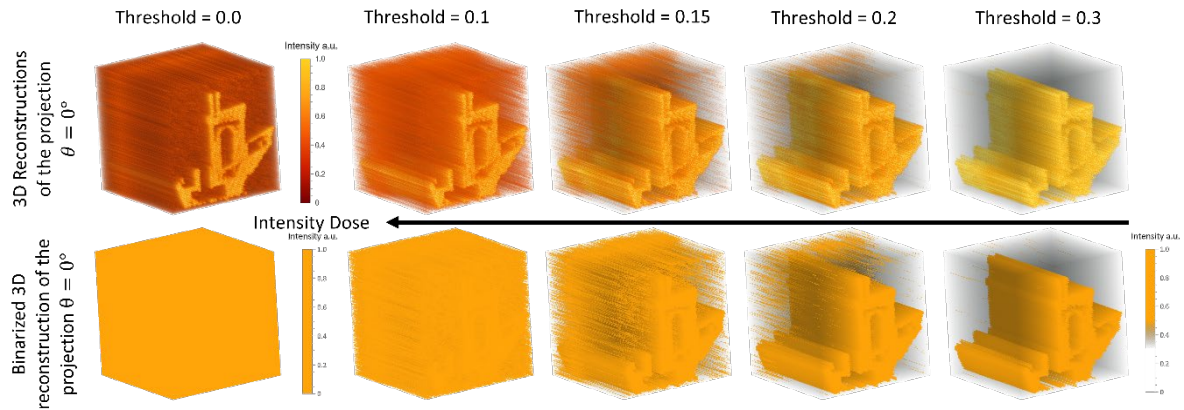

**Fig 7**. Effect of dose threshold in 3D printing on speckle noise. Simulation of a 3D reconstruction corresponding to the projection angle $\theta = 0°$. Different thresholds where applied (top) to simulate the intensity dose applied to the resin. The correct applied dose avoids the printed part to print the speckle noise. Binarized 3D reconstructions of the applied dose (bottom).

## *Volumetric additive manufacturing with tomographic holographic projections*

We used the holographic projectiosn to fabricate millimetric objects within seconds. Here, a commercial polyacrylate resin with Diphenyl (2,4,6-trimethylbenzoyl) phosphine oxide (TPO) as a photoinitiator, was solidified to a desired shape using the computer-generated holograms. We demonstrated the ability to fabricate millimetric 3D objects in less than a minute printing a model of the 3D Benchy boat, as shown in Fig 8. SEM micrography and micro-CT scans of the printed boats show our ability to fabricate small details, such as the sharp bow of the boat or the hollow cabin (Fig. 8.B-C). We realized minimal positive features of 164 µm and minimal negative features of 240 µm (Fig. 8. D).

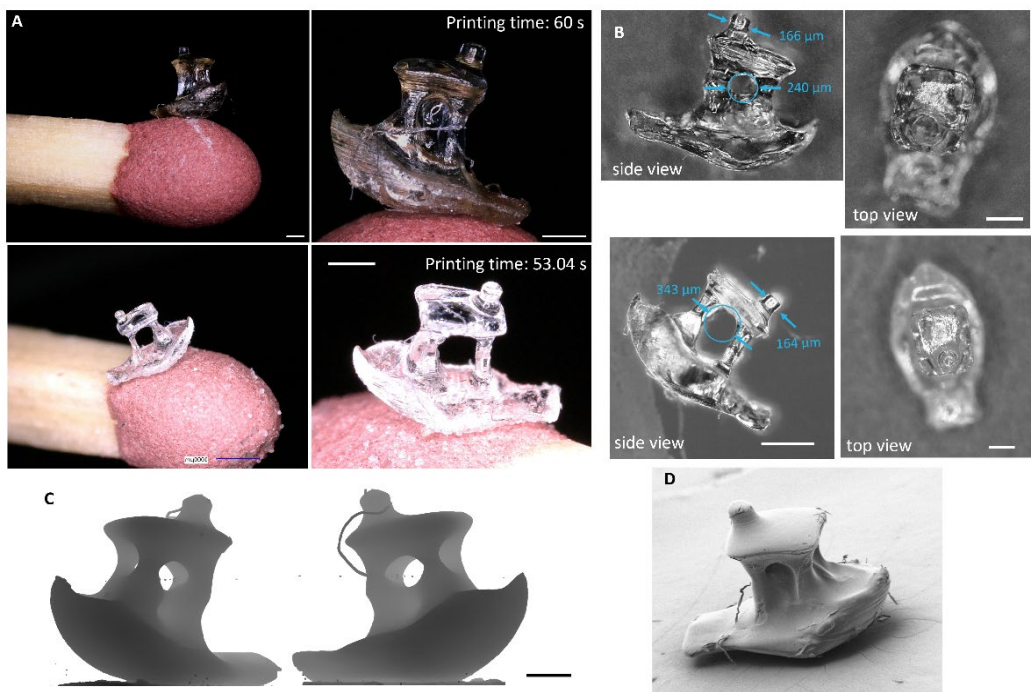

**Fig 8. Exemplary prints of the 3D Benchy boat fabricated in ≥60 s. (A)** Micrographs of the printed part sitting on the head of a match.. **(B)** Side views from micro-CT reconstruction. **(C)** Printed parts exhibit features below 200 µm. Scale bars 500 µm **(D)** SEM images.

## IV.     Discussion and Conclusion

In this work we have demonstrated the use of holographic phase encoding for tomographic volumetric additive manufacturing. We show that by rapidly projecting holograms onto the rotating photoresin, we can produce synthetic low-divergence light beams. These low-divergence beams match more closely the collimation assumption of the Radon transform, the computational backbone to calculate tomographic projections. We illustrate the low-divergence of these phase-encoded projections through computational simulations and experimental recordings.

Light intensity efficiency is key to fabricating objects within seconds in tomographic VAM approaches. This is due to the low concentration of photoinitiators. An additional advantage of using phase-encoding instead of only amplitude encoding is that the efficiency of the spatial light modulator increases dramatically. In amplitude projections, very few pixels on the digital micromirror device contribute to the image, wasting most of the light. Alternatively, in phase-encoding, most pixels are in an ON-state. Thanks to this, we could realize additive manufacturing of millimetric objects with a 40 mW cw laser source, instead of the 1-3 W input powers typically used in other works [8]. The optomechanics of the required light engine are also simpler and could compensate for downstream optical aberrations.

We also introduced a calculation pipeline to compute holograms for tomographic projection. A disadvantage of phase-encoding and holographic projection is speckle noise. Noisy projections may result in prints of low fidelity. Thus, we used the *speckle-reduction* method called HoloTile, for digital holography [42], [44], to compute tiled holograms with superposed point spread function modification. The addition of Fresnel lenses allowed us to produce a sequence of holograms to obtain *low etendue* projections. In addition, we use *Lee Holograms* [45] in TVAM systems to convert the binary amplitude DMD as a fast phase modulator. This method may open the possibility of adapting current light-based 3D printing systems based on incoherent light engines to coherent light engines that offer much higher efficiency and flexibility in the light control.

These results are the first demonstration of holographic and tomographic phase-encoded light projection for volumetric additive manufacturing. They are an addition to the family of vat-photopolymerization methods, and may open new avenues for additive manufacturing at lower powers, better light control and with simplified light engines.

## V.     Materials and Methods

*Optical setup*

Figure 1 shows the optical configuration used to produce holographic projections in a volumetric printer. It consists of one 40 mW CW laser diode at 405nm (OBIS 405nm LX SF 40mW) coupled into a single-mode fiber (Thorlabs P3-405B-FC) to obtain a good quality Gaussian beam. A collimated beam illuminates a DMD (Vialux DLP7000: $L \times L$ =768 × 1024 micro-mirrors and pixel pitch = 13.6 μm ) placed in a Fourier configuration and aligned to meet the blazed grating criteria of efficiency. The DMD was synchronized with the rotary using a Data Acquisition Card from National Instruments at a typical framerate of 1600 Hz (32 holograms per $\Delta\theta$, $\Delta\theta$ =0.6°, 12 s / rotation.).

After the Fourier Lens L1 ($f_1$ = 150mm), a spatial filter (SF) was added to filter out the zero-order and keep the +1 order. L2 ($f_2$ =150 mm) and L3 ($f_3$ = 200 mm) form a 4-*f* system conjugating the Fourier Plane and rescaling the holograms on the sample plane. Two inspection systems using Lens L4 ($f_4$ = 100 mm) with camera 1 (iDS UI307xCP-M), and Lens L5 ($f_5$ = 30 mm) with camera 2 (iDs UI327xCP), were implemented for the inspection of the holographic projections, and polymerization process respectively. The schematic of the experimental setup is shown in the Supplementary Material S-Figure 1.

*Photocurable resins*

Resins were prepared by mixing the photoinitiator TPO (Diphenyl (2,4,6-trimethylbenzoyl)- phosphine oxide, Sigma) in a polyacrylate commercial resin (PRO 21905, Sartomer) to a concentration of 3 mM using a planetary mixer (Kurabo Mazerustar). The photocurable resin was poured into cylindrical glass vials (outer diameter 12 mm), and then sonicated to remove air bubbles.

*Post-processing of printed parts*

Printed parts were recovered from the glass cylinders and rinsed for 10 minutes in propylene glycol monomethyl ether acetate (PMGEA) with slow agitation in a vortex mixer. Then, where cleaned in isopropyl alcohol (IPA) for 10 minutes more. Finally, they were post cured under UV light while immersed in PMGEA.

*Imaging*

Micro-CT scans

Printed objects were imaged with voxel sizes of $10 \times 10 \times 10$ µm$^3$ or $2 \times 2 \times 2$ µm$^3$ under a 160 kV X-ray transmission tomograph (Hamamatsu, Japan). 3D visualizations and cross sections of the pieces were obtained using Fiji-ImageJ[52].

*Divergence of projections*

To record the 3D projections of the CGH in the build volume, camera 2 (UI327xCP, IDS Imaging) in the setup was mounted on a linear motorized micrometric stage (Z912B, Thorlabs). The servo motor was controlled and synchronize with the camera acquisition so that the camera would record a photograph every 20 µm.

*Photography*

Printed parts were imaged with a DSLR camera (D3100, Nikon) with a $f$ = 2.8 macro lens (AF-S Micro Nikkor 40 mm, Nikon), and a digital microscope (VHX-5000, Keyence) with magnifications between 20 and 200x.

*Simulations*

3D renderings of .stl files. Recorded light intensities, or simulates intensity distributions were produced with Wolfram Mathematica® 13.1 [53]. To calculate the intensity projections using the Radon transform, the CGHs, and simulations of propagations using Angular Spectrum (AS) were done using MATLAB® [54] and Wolfram Mathematica® 13.1[53].

*3D models*

The 3D .stl models of the Benchy Boat (https://free3d.com/de/3d-model/3dbenchy-the-jolly-3d-printing-torture-test-54655.html; licensed under CC BY-ND 4.0) and a custom-made gear were used.

*Hardware control*

To drive and synchronize the devices in the implemented printed, we used Python. To control the DMD we used the ALP4lib Module of Sébastien M. Popoff [55]. For the rotatory motor stage we used the Zaber library (https://www.zaber.com/software/docs/motionlibrary/ascii/references/python/). We used a National Instruments DAQ card (PCIe-6321, X Series) as a reference clock to trigger the sequence display on the DMD and the camera 1 and 2.

*Acknowledgements*

This project has received funding from the Eurostars-3 (VOLTA-E!3908) joint programme with co-funding from the European Union's Horizon Europe research and innovation programme, Innosuisse (Swiss Innovation Agency) and Innovation Fund Denmark (IFD), and from the Swiss National Science Foundation under project number 196971 - "Light based Volumetric printing in scattering resins."

The authors would like to acknowledge Gary Perrenoud, Lionel Pittet, and Albert Taureg (PIXE Platform, EPFL) for their support with microCT imaging of the printed structures. The authors thank Buse Ünlü for her help with SEM imaging at the Center of MicroNanoTechnology (EPFL).

# Holographic Volumetric Additive Manufacturing: Supplementary Material

*Maria I. Álvarez-Castaño[1], Andreas Gejl Madsen[2], Jorge Madrid-Wolff[1], Antoine Boniface[1,3], Jesper Glückstad[2], Christophe Moser[1]*

[1]Laboratory of Applied Photonics Devices, School of Engineering, Ecole Polytechnique Fédérale de Lausanne, CH-1015, Lausanne, Switzerland
[2]SDU Centre for Photonics Engineering, University of Southern Denmark, DK-5230 Odense M, Denmark
[3] current address: AMS Osram, Martigny, Switzerland

## S1. Experimental setup

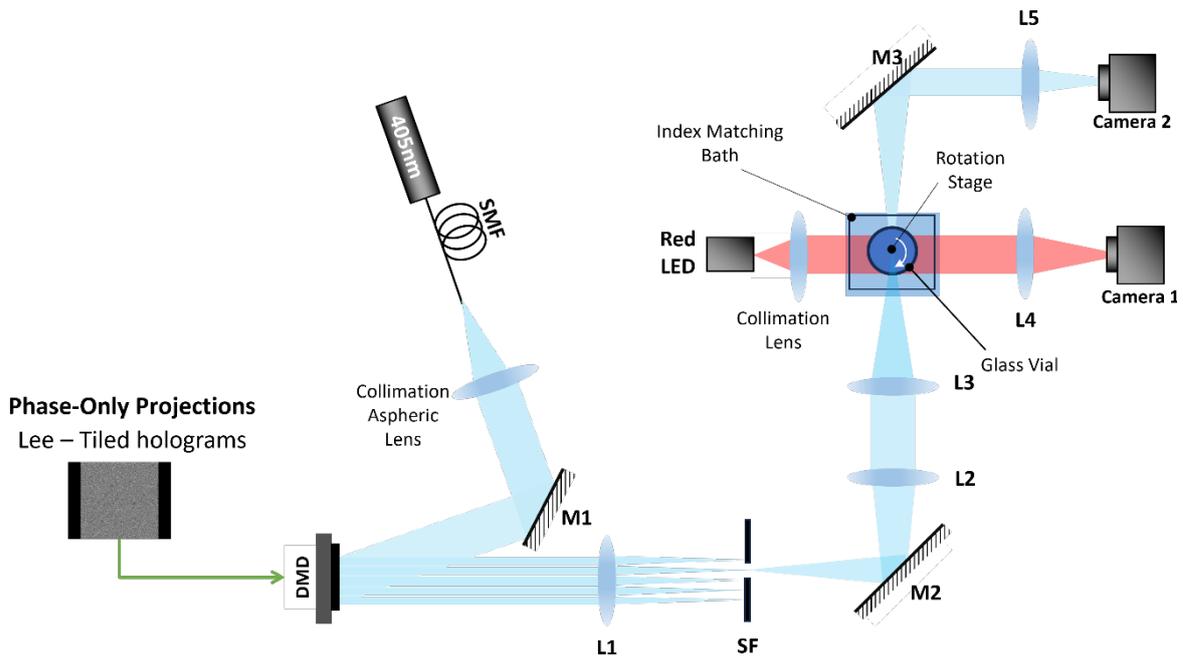

**S-Figure 1.** Experimental setup of the Tomographic Volumetric 3D printing by phase encoding

## S2. Hologram test: Speckle noise

To compare the intensity reconstruction provided by a LC-SLM and DMD using Lee Holograms Method, and the effect of time multiplexing. We generated a CGH using the letter "A" as a target intensity to display on LC-SLM, and a a CGH of the letter "E" encoded by Lee Hologram to display on the DMD. S-Figure 2 (B) and (D) shows how multiplexing on time six holograms the contrast of the reconstructions increases indistinguishing the hologram projector used, DMD or LC-SLM.

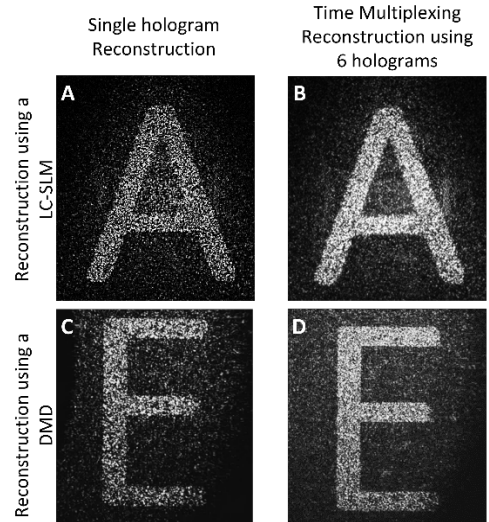

**S-Figure 2.** CGH test. (**A**) Intensity reconstruction of a single CGH with letter "A" as a target intensity. The CHG was display in a LC-SLM using the experimental setup in figure S-Figure 3. (**B**) Image of an intensity reconstruction of six time multiplexed holograms of the letter "A" using a LC-SLM. Time multiplexing increase the contrast. (**C**) Intensity reconstruction of a single CGH with letter "E" as a target intensity. The CHG was display in a DMD thanks to Lee Hologram Method. (**D**) Image of an intensity reconstruction of six time multiplexed holograms of the letter "E" letter using a DMD thanks to Lee Hologram Method.

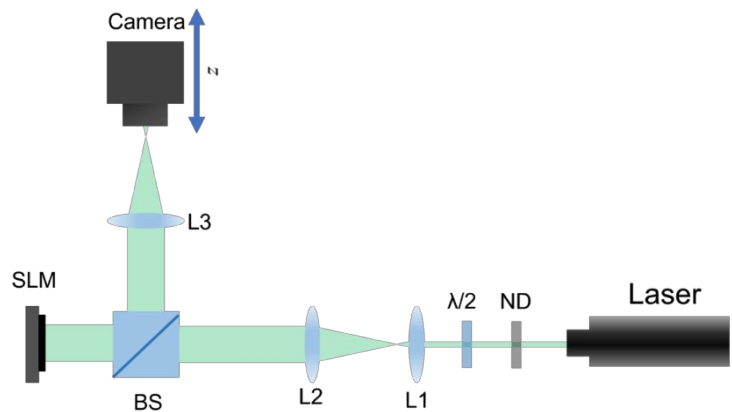

**S-Figure 3. Experimental setup to test CGHs using LC-SLM.** A CW laser at 532 nm was used as light source. Lenses $L1$ and $L2$ were used to expand the beam by a factor of 5. A liquid crystal SLM (Holoeye PLUTO NIR II) was used to display the CGH, with a $\lambda2/$ waveplate used to provide the necessary polarization to obtain phase-only modulation. Normal incidence on the SLM was chosen to avoid additional aberrations in the reconstruction. The reconstructed images were captured using a monochromatic CMOS camera with a pixel size of 2.2 µm.

### S3. Light efficiency measurements

The same intensity projected pattern in S-Figure 4A can be generated using phase coding instead of amplitude coding. In this case, all pixels in the modulator plane contributes to one pixel in the image plane. This is illustrated in S-Figure 4B, in which the image plane is in the Fourier plane of the lens. To measure the efficiency of both light engine we use the experimental setup shown in S-Figure1. Here, first we measured the power when all the pixels are "ON" after the spatial filter (SF) $P_{in}$ = 4.61 mW. Then, the projection pattern shown in (A) is projected in the DMD. The mirror M2 is placed in a flip mount adapter (FM90, Thorlabs) that allows the light propagates straight, a lens L2 (f₂ =150 mm) is placed in the path and the power of the projected pattern is measure in the image plane $P_{out}^{incoherent} = 0.016\ mW$, providing an efficiency of the incoherent patterns of $\eta_{incoherent} = 0.34\%$. For coherent pattern instead, the power in the Fourier plane after the $4f$ system is measured when the Lee

Hologram corresponding to the projection $\theta = 0°$ is displayed on the DMD. The power in the Fourier plane is $P_{out}^{coherent} = 0.260\ mW$, giving an efficiency of $\eta_{coherent} = 5.6\ \%$. The light efficiency here could increase even more if we use the full size of the DMD display, now we are using an hologram with size 768 x 768 pixels due to the constrains of the PSF shaping.

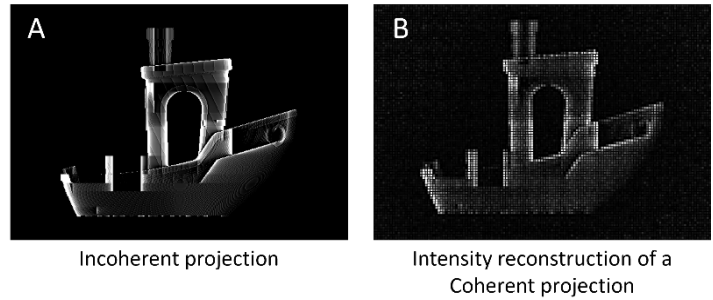

Incoherent projection    Intensity reconstruction of a Coherent projection

**S-Figure 4.** (**A**) Amplitude pattern corresponding to the projection angle $\theta = 0°$. (**B**) Intensity reconstruction of coherent projections generated using as a target intensity the amplitude pattern in (A).

### S4. Sampling Rate and Resolution

Since the system is diffraction limited, the voxel resolution is given by $\propto \lambda/NA$. The NA is calculated considering the sub-hologram size as the aperture size that changes with the number of tiles, and the magnification of the 4f system. We can estimate that for tiles $N_t < 5$ the resolutions are in between $1.5\mu m < \lambda/NA < 9.2\ \mu m$ for a magnification of 1.3x, we can see from S-Figure. 5A. In our system, the holograms synthesis was done considering tiled holograms with $N_t = 4$ , due to it is possible to reach a speckle noise reduction with a contrast $\sim 0.9$ and a voxel resolution of $\sim 6.2\ \mu m$. The calculation are considering a full matrix of 768 X 768 pixels with a pixel size of 13.7 $\mu m$, and $\lambda = 405\ nm$.

S-Figure 5B. shows the sample rate calculations in terms of the number of tiles, here we is possible to see how the number of "positions" or addressable points in the reconstruction plane decreases with the number of tiles.

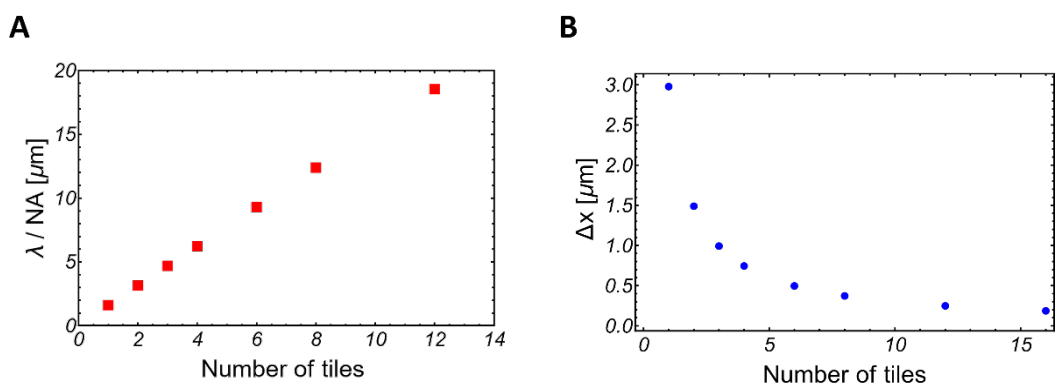

**S-Figure 5.** (**A**) Resolution calculation for tiled holograms as $\lambda/NA$ (**B**) Sampling rate calculation for different tiled holograms

### S5. Gerchberg-Saxton iterative synthesis of holograms

Computer-generated holograms (CGH) were synthesized using the Gerchberg-Saxton (GS) algorithm [1], [2] to generate the desired intensity pattern in the sample plane by modulating the phase of the input wavefront of a coherent light beam. Here, the tomographic projections previously computed using the filtered back projections were used as target images. To retrieve a 2D phase mask, the algorithm performs *m* iterations by simulating forward and backward propagation of the light field from the sample plane to the Fourier plane through a direct inverse fast Fourier Transform (S-Figure 6.A-B). This is performed on every projection.

S-Figure 6.C illustrates the block diagram of the traditional GS algorithm. In the first iteration, an initial guess phase $\Phi_0(x, y)$, in our case an output phase from a superposition GS, where a random phase is use as a input. Then, this phase is applied to the target amplitude at the image plane $\sqrt{I_T} \times \exp[j \times \Phi_0]$. The field is inverse Fourier transformed to the hologram plane, where the amplitude $u$ is discarded and replaced by the amplitude of the illumination source $\sqrt{I_s} \times \exp[j \times \varphi]$. The new complex field is Fourier transformed again to the image plane. The resulting amplitude $U$ is replaced by the target intensity and the newly computed phase $\Phi(x, y)$ is kept. After performing $m$ iterations, the final phase $\Phi_m(x, y)$ is used as the phase CGH [1].

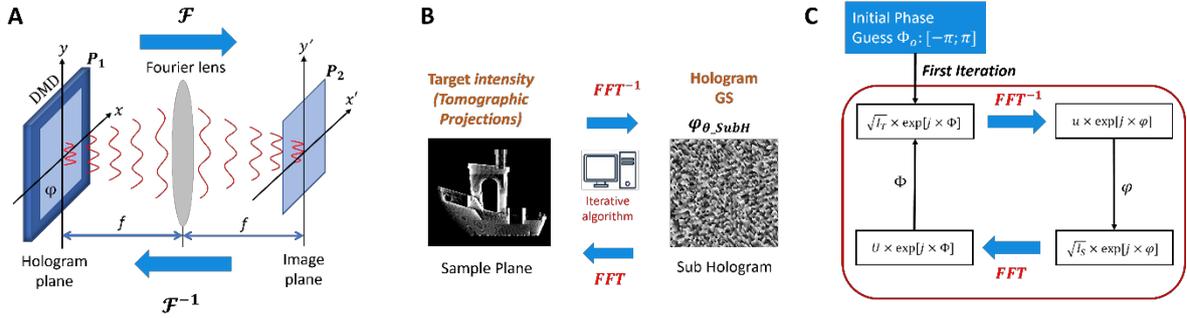

**S-Figure 6.** (**A**) Schematic diagram of the optical Fourier transform system to illustrate the forward and backward propagation of the light in a GS algorithm. (**B**) Block diagram of the forward and backward propagation of a GS algorithm using Tomographic projections as a target intensities. (**C**) Block diagram of a traditional GS algorithm.

### *S6. General HoloTile formalism*

Using Fourier analysis on the applied Fourier holographic *2f*-setup we can perform a semi-analytical assessment of the inner workings of the HoloTile Light Engine [3] using either a complex (amplitude and phase or phase-only or amplitude-only) spatial light modulator at the input. Assuming a photon-efficient *phase-only* input displayed at an SLM (or in-directly from a DMD using Lee holograms [4], [5]) the read-out input light field can be described as:

$$a = b(x,y)\text{rect}\left(\frac{x}{X_{SLM}}, \frac{y}{Y_{SLM}}\right)\exp(i\varphi_{PSF}(x,y))\left[\exp(ih(x,y)) \otimes \text{comb}(x\Delta X, y\Delta Y)\right] \quad (1)$$

where $b(x,y)$ correspond to the input beam, $\text{rect}\left(\frac{x}{X_{SLM}}, \frac{y}{Y_{SLM}}\right)$ corresponding to the pixeld structure of the display, $h(x,y)$ is the calculated CGH to be tiled, and $\varphi_{PSF}(x,y)$ is an independent PSF-shaping hologram phase. The tiled hologram is $\varphi_{tile}(x,y) = \left[\exp(ih(x,y)) \otimes \text{comb}(x\Delta X, y\Delta Y)\right]$.

The resulting complex field reconstruction at the optical Fourier plane of the 2*f*-setup:

$$\tilde{O} = \frac{X_{SLM}Y_{SLM}}{\Delta X \Delta Y} \text{sinc}(X_{SLM}f_x, Y_{SLM}f_y) \otimes \Im(b(x,y)) \otimes \Im(\exp(i\varphi_{PSF}(x,y)))$$
$$\otimes \left[\Im\left(\exp(ih(x,y))\right) \text{comb}\left(\frac{f_x}{\Delta X}, \frac{f_y}{\Delta Y}\right)\right] \quad (2)$$

Assuming a plane wave read-out we have $\Im(b(x,y)) \to \delta(0,0)$ and therefore the simpler expression:

$$\tilde{O}_{\rightrightarrows} = \frac{X_{SLM}Y_{SLM}}{\Delta X \Delta Y} \text{sinc}(X_{SLM}f_x, Y_{SLM}f_y) \otimes \Im(\exp(i\varphi_{PSF}(x,y)))$$
$$\otimes \left[\Im\left(\exp(ih(x,y))\right) \text{comb}\left(\frac{f_x}{\Delta X}, \frac{f_y}{\Delta Y}\right)\right] \quad (3)$$

We can now better analyze the interplay between the HoloTile reconstructed output terms.

The optical Fourier reconstructed output term arising from the tiled hologram $h(x,y)$ is given by:

$$\tilde{O}_H = \Im\left(\exp(ih(x,y))\right) \text{comb}\left(\frac{f_x}{\Delta X}, \frac{f_y}{\Delta Y}\right) \quad (4)$$

which basically describes a *sampled* reconstruction of the Fourier transformed hologram phase. The output term arising from the independent PSF-shaping hologram phase is given by:

$$\tilde{O}_P = \Im\left(\exp(i\varphi_{PSF}(x,y))\right) \quad (5)$$

By ignoring the diffraction impact from the input SLM/DMD aperture truncation – corresponding to approximating the term $\text{sinc}(X_{SLM}f_x, Y_{SLM}f_y) \to \partial(0,0)$ with an output centered Kronecker delta-function for increasing aperture sizes - we finally have the following simple expression for the HoloTile Light Engine reconstructed complex output field:

$$\tilde{O}_{\rightrightarrows} \propto \tilde{O}_P \otimes \tilde{O}_H \quad (6)$$

Let us first consider the optical Fourier reconstructed output term, $\tilde{O}_H$, arising from the tiled hologram $h$ displayed at the input spatial light modulator or DMD:

$$\tilde{O}_H = \Im\left(\exp(ih(x,y))\right) \text{comb}\left(\frac{f_x}{\Delta X}, \frac{f_y}{\Delta Y}\right) \quad (7)$$

This term can be illustrated as a delta-function discretized and amplitude weighted version of the "continuous" holographic reconstruction $\Im\left(\exp(ih(x,y))\right)$.

2D-convolved onto this delta-function discretized and amplitude weighted holographic reconstruction we have the output term arising from the PSF-shaping hologram phase as given by:

$$\tilde{O}_P = \Im\left(\exp(i\varphi_{PSF}(x,y))\right) \tag{8}$$

The resulting effect can be illustrated by superposing $\Im\left(\exp(i\varphi_{PSF}(x,y))\right)$ around each amplitude weighted delta-function discretization generated by $o_H$. Examples are illustrated below in space and/or time. The same applies depth-wise [6] along the optical axis $z$ so that we can also write:

$$\tilde{O}_\rightrightarrows(f_x, f_y, \Delta z, t) \propto \widetilde{O_P}(f_x, f_y, \Delta z, t) \otimes \tilde{O}_H(f_x, f_y, t) \tag{9}$$

*S6.1. Point spread function modification*

To calculate the PSF phase $\varphi_{PSF}(x,y)$ to convert a single-mode Gaussian beam into a flat-top, the parameter $\beta$, which is a dimensionless parameter, is important to set the system to do a relatively good beam shape at the target plane [7],[8] . Higher values of $\beta$ are related with a good geometrical approximation to a flat-top.

$$\beta = \frac{2\sqrt{2\pi}r_0 y_0}{f\lambda} \tag{10}$$

Here $\lambda$ is the wavelength, $r_0$ is the radius at $1/e^2$ point of the input beam, $y_0$ is half-width of the desired dimension in the reconstruction plane, and $f$ is the focal length of the Fourier lens [7], [8]. The $\varphi_{PSF}(x,y)$ is dependent on $\beta$ and the phase element $\varphi(\xi)$. The analytical function to shape the flat-top is

$$\varphi_{PSF}(x,y) = [\beta_x \varphi_x(x) + \beta_y \varphi_y(y)] \tag{11}$$

It is necessary to calculate $\beta$ and $\varphi(\xi)$ for each dimension.

Where, $\varphi(\xi) = \frac{\sqrt{\pi}}{2} \xi\, \text{erf}(\xi) + \frac{1}{2}\exp(-\xi^2) - \frac{1}{2}$, and $\xi = \frac{\sqrt{2}\cdot x}{r_0}$ or $\xi = \frac{\sqrt{2}\cdot y}{r_0}$ [3],[4].

The relationship between the Fourier plane and the DMD plane (aperture) is given by the relation [9]:

$$\Delta x = \frac{\lambda f}{w_x} \tag{12}$$

Where $\lambda$ is the wavelength of the light, $f$ is the focal length of the Fourier lens, $w_x$ is the half-width of the aperture, and $\Delta x$ is the half width of the period of the interference pattern at the Fourier plane. Considering the physical parameters related to the DMD and sub-holograms we can rewrite the equation as:

$$\Delta x = \frac{N_t \lambda f}{2L \cdot \ell_{px}} \tag{13}$$

Where $\ell_{px}$ is the pixel size in micrometers. From equation (13) it is possible to see that for each tile there is a modification of the sampling function which is related to the half-width of the desired flat-top dimension in the Fourier plane, as the number of tiles modifies the separation between the image points. The flat-top generation is highly dependent from the sampling rate [7], and is giving by an analytical function see Supplementary Material Point spread function modification.

After adding a phase function to provide a flat top reconstructed pixel, the final phase of the tiled CGH encoding by HoloTile is:

$$\varphi_{Tiled} = \varphi_{tile} + \varphi_{PSF} \tag{14}$$

To calculate the phase encoding to produce the flat-top, we insert equation (13) into equation (7) and set $y_0 = \Delta x$ from equation (13).

*S7. Lee Hologram Method*

Binary-amplitude holography enabled fast wavefront control using a DMD to display the phase-encoded tomographic projections. Lee Holograms is a simple technique to modulate the amplitude $A(x,y)$ and phase $\varphi(x,y)$ of a target optical field. The binarization of the hologram is easy to produce following equation (15) [4], [5]:

$$H(x,y) = \frac{1}{2} + \frac{1}{2} sgn[\cos[2\pi v_0 x + \varphi(x,y)] - \cos[\pi\omega(x,y)]] \tag{15}$$

Here the phase is encoded using a linear carrier in $x$ with frequency $v_0$, $sgn$ is the sign function, and $\omega(x,y)$ is a function related with the amplitude of the target optical field as follows:

$$\omega(x,y) = \frac{1}{\pi} \arcsin[A(x,y)] \tag{16}$$

The linear carrier with frequency $v_0 = 1/x_0$, separate the diffraction orders in the Fourier plane. Equation (15) is simplified for a phase-only hologram by considering a uniform and unitary amplitude. The axial control is achieved by superposing with a Diffractive Optical Element (DOE), in this case a Fresnel lens, with the designed holograms before the encoding to a binary hologram. Mathematically the phase lens [10],[11] used to the divergence control of the CGH is expressed as:

$$\varphi_{Len}(x,y,z) = \frac{2\pi z}{\lambda f^2}(x^2 + y^2) \tag{17}$$

Where $z$ is the desired shift of the reconstruction plane relative to the focal length $f$ of the Fourier lens. Adding the Fresnel lens phase $\varphi_{Len}(x,y,z_n)$ to a tiled phase $\varphi_{Tiled}(x,y)$ gives the effect of moving the reconstruction plane of the tiled hologram a distance $z_n$ around the focal plane. The effect of tiling the holograms and the flat-top phase determined by equation (14) is also shifted by the Fresnel phase. A stack of $N$ holograms per projection angle was used to spread the tomographic projections with a controllable divergence over the propagation axes, each phase is calculated from equation (18).

$$\varphi_\theta^N(x,y,z_n) = \varphi_{Tiled}(x,y) + \varphi_{Len}^N(x,y,z_n) \tag{18}$$

Each phase set $\varphi_\theta^N$ is encoded in a binary Lee Hologram. Where the Lee holograms sequences displayed on the DMD is given by:

$$H_\theta^N(x,y,z_n) = \frac{1}{2} + \frac{1}{2} sgn\left[\cos\left[\frac{2\pi x}{x_0} + \varphi_\theta^N(x,y,z_n)\right]\right] \tag{19}$$